\documentclass[superscriptaddress,amsmath,amssymb]{revtex4}
\usepackage{color}
\usepackage{graphicx}
\usepackage{amsmath}
\usepackage[utf8x]{inputenc}
\usepackage[T1]{fontenc}

\begin{document}
\title{Rotation-time symmetry in bosonic systems and the existence of exceptional points in the absence of $\mathcal{PT}$ symmetry}

\author {E. Lange}
\affiliation{Faculty of Physics, Adam Mickiewicz University, 61-614 Pozna\'{n}, Poland}
\author {G. Chimczak}
\email{chimczak@amu.edu.pl}
\affiliation{Faculty of Physics, Adam Mickiewicz University, 61-614 Pozna\'{n}, Poland}
\author {A. Kowalewska-Kud{\l}aszyk}
\affiliation{Faculty of Physics, Adam Mickiewicz University, 61-614 Pozna\'{n}, Poland} 
\author {K. Bartkiewicz}
\affiliation{Faculty of Physics, Adam Mickiewicz University, 61-614 Pozna\'{n}, Poland}
\affiliation{RCPTM, Joint Laboratory of Optics of Palacký University and Institute of Physics of Czech Academy of Sciences, 17. listopadu 12, 771 46 Olomouc, Czech Republic}

\begin{abstract}
We study symmetries of open bosonic systems in the presence of laser pumping. 
Non-Hermitian Hamiltonians describing these systems can be parity-time (${\cal{PT}}$) symmetric in special cases only. Systems exhibiting this symmetry are characterised by real-valued energy spectra and can display exceptional points, where a symmetry-breaking transition occurs. We demonstrate that there is a more general type of symmetry, i.e., rotation-time (${\cal{RT}}$) symmetry. We observe that ${\cal{RT}}$-symmetric non-Hermitian Hamiltonians exhibit real-valued energy spectra which can be made singular by symmetry breaking. To calculate the spectra of the studied bosonic non-diagonalisable Hamiltonians we apply diagonalisation methods based on bosonic algebra. Finally, we list a versatile set rules allowing to immediately identifying or constructing ${\cal{RT}}$-symmetric Hamiltonians.  We believe that our results on the ${\cal{RT}}$-symmetric class of bosonic systems and their spectral singularities can lead to new applications inspired by those of the ${\cal{PT}}$-symmetric systems.
\end{abstract}

\maketitle

\section{Introduction}
Symmetries are among the most fundamental concepts in physics. They can be viewed as the source of the conservation laws by Noether's theorem. They also result in degeneracies. For example, degeneracies of open, non-conservative systems, described by non-Hermitian Hamiltonians, can be associated with symmetry breaking. A special type of a spectral singularity in the parameter space of such systems, where two or more eigenvalues coalesce, is known as an exceptional point (EP)~\cite{Ozdemir2019,Miri2019,el2018non}. EPs have been usually studied in the context of $\mathcal{PT}$-symmetric non-conservative semi-classical systems~\cite{bender1998real,Christodoulides:2018arz,PengScience14}. However, EPs can be studied in systems without any relation to the $\mathcal{PT}$ symmetry~\cite{Ozdemir2019,Miri2019}. Recently, the concept of EPs have been generalised to describe fully quantum systems including quantum jumps~\cite{PhysRevA.100.062131,PhysRevA.101.013812,minganti2020hybridliouvillian}. Possible applications of EPs for quantum sensing have been attracting much interest (see~\cite{Lau2018} and references therein). EPs are widely investigated particularly in optics, where dissipative, non-Hermitian systems commonly appear. However, the intriguing physics of symmetry-breaking transitions recently has been attracting attention of scientists in other fields like acoustics \cite{shi2016accessing, PhysRevB.95.144303,PhysRevX.6.021007}, atomic physics \cite{PhysRevLett.99.173003}, and photonic crystals \cite{zhen2015spawning, PhysRevLett.117.107402}. The interplay between gain and loss, an intrinsic aspect of open systems, leads to new, fascinating effects. The examples of recently discovered phenomena associated with EPs include exceptional photon blockade \cite{huang2020exceptional}, unidirectional light propagation \cite{yin2013unidirectional}, directional lasing \cite{peng2016chiral}, topological energy transfer \cite{xu2016topological}, and
other phenomena~\cite{zhong2018power, zhang2019quantum, chen2017exceptional, PhysRevLett.106.213901, feng2014single, peng2014parity, feng2013experimental, hodaei2014parity, chang2014parity, fleury2015invisible, PhysRevLett.112.143903}. Not only the exact point of symmetry breaking leads to remarkable physics, but also the systems found in the vicinity of EPs can exhibit features as potentially enhanced sensitivity~\cite{PhysRevLett.117.110802, hodaei2017enhanced}. The latest studies are commonly associating real-valued spectra with the respective $\cal{PT}$-symmetry requirements and EPs with $\cal{PT}$-symmetry breaking~\cite{bender1998real, bender2007making, BenderBook, el2018non, garcia2017spectrum, Nixon2015, ozdemir2019parity}.

In this paper, we introduce the notion of rotation-time ($\cal{RT}$) symmetric bosonic systems, where we demonstrate coalescence of energy values. The concept of rotation-time symmetry was previously introduced only in fermionic systems~\cite{zhang2013non, PhysRevA.88.042108, wang2020effective}. However, bosonic systems are currently a topic of intense research~\cite{garcia2017spectrum, ohashiPRA, chimczak2018creating, zhong2018power, AnnaGrzegorz2019, Zhang2019higher,Perina2019,HeuckPRL20} due to being a promising platform for gain and loss engineering in physical experiments~\cite{el2018non}. We demonstrate that $\cal{RT}$ invariance allows a given system to have a real energy spectrum, which becomes singular, as a result of a $\cal{RT}$-symmetry phase transition. Furthermore, we demonstrate that the $\cal{RT}$ symmetry is a superset of the $\cal{PT}$ symmetry, i.e, we identify a wide class of Hamiltonians that have similar properties to those of the $\cal{PT}$-symmetric systems, while not necessarily being $\cal{PT}$-symmetric.

We also find general rules for constructing $\cal{RT}$-symmetric symmetric non-Hermitian Hamiltonians. This provides a framework for exploring the physics of singular energy spectra in terms of symmetries in a range of bosonic systems including non-linear interactions between modes, associated, e.g., with unconventional photon blockades or second- and third-harmonic generation processes~\cite{HeuckPRL20,FHPW61,B08,S84,KMMSN17}.
Consequently, future experiments could explore properties of spectral singularities in areas of quantum and atom optics that have never been studied before.

\section{Results}

\subsection{${\cal{PT}}$ symmetry in bosonic system}
Let us begin by considering a ${\cal{PT}}$-symmetric Hamiltonian describing a damped bosonic linear system with a classical laser drive. 
\begin{figure*}[ht]
    \centering
    \includegraphics[width=0.6\linewidth]{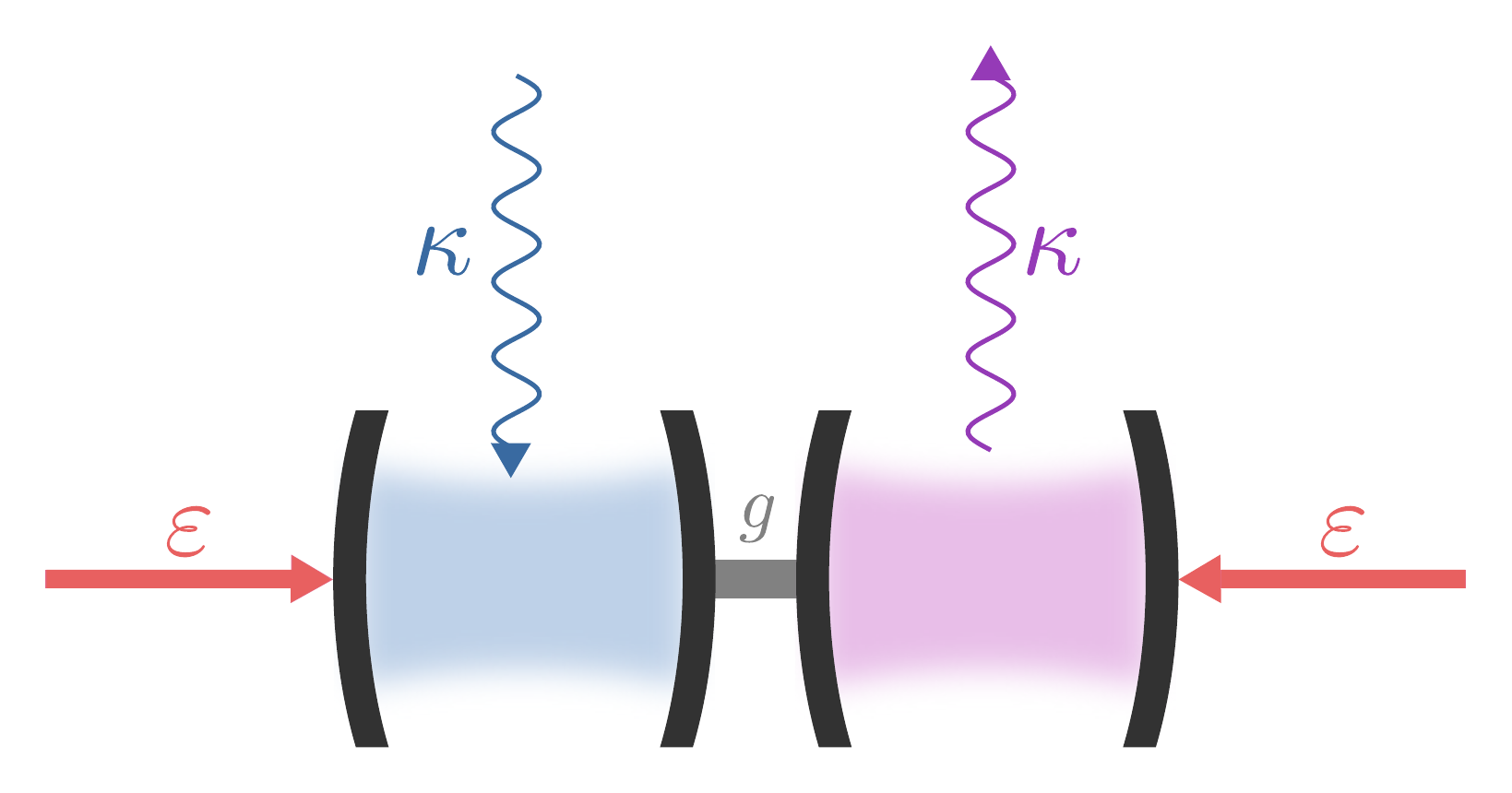}
    \caption{The scheme of the setup we consider to observe the rotation-time symmetry. It consists of two linearly-coupled cavities (one with loss and another with gain) and two lasers driving both cavities.}
    \label{fig:model}
\end{figure*}
Thus, let us choose a system depicted in Fig.~\ref{fig:model} and described with Hamiltonian
\begin{eqnarray}
  \label{eq:HamiltonianPT}
  H_{1}&=& g (a^{\dagger}b + b^{\dagger}a)
    +\varepsilon (i a - i a^{\dagger}) +\varepsilon (i b - i b^{\dagger})
    -i\kappa a^{\dagger}a+i\kappa b^{\dagger}b    \, ,
\end{eqnarray}
where $a$ and $b$ are the annihilation operators for two modes, respectively,
$g$ is the coupling strength between these two modes, $\kappa \equiv \kappa_a = \kappa_b$ is the field loss/gain rate of the $a/b$ mode and $\varepsilon$ is the strength of the modes drive. We assume for the sake of simplicity that $g$ and $\varepsilon$ are real and positive. It is known that a Hamiltonian $H$ is regarded as ${\cal{PT}}$-symmetric if it commutes with the operator ${\cal{PT}}$.
Since ${\cal{PT}}$ is a reflection, i.e., ${\cal{PT}}=({\cal{PT}})^{-1}$, we can rewrite the condition for $H$ to be ${\cal{PT}}$-symmetric as $({\cal{PT}})H({\cal{PT}}) = H$.
So, if the transformation does not change $H$, then $H$ is ${\cal{PT}}$-symmetric.
Now we want to check whether $H_{1}$ has ${\cal{PT}}$-symmetry properties. For that purpose, it is convenient to know how ${\cal{PT}}$ transforms bosonic operators.
For a single field mode, one can easily derive all needed formulas knowing the effect of the space-reflection operator (also known as the parity operator) ${\cal{P}}$ and the time-reversal operator ${\cal{T}}$, where  $[{\cal{P}},{\cal{T}}]=0$, on the position operator $\hat{x}$ and the momentum operator $\hat{p}$: ${\cal{P}}\hat{x}{\cal{P}}=-\hat{x}$, ${\cal{T}}\hat{x}{\cal{T}}=\hat{x}$, ${\cal{P}}\hat{p}{\cal{P}}=-\hat{p}$, and ${\cal{T}}\hat{p}{\cal{T}}=-\hat{p}$.
From the form of the position operator $\hat{x}=(a+a^{\dagger})/\sqrt{2}$ and the momentum operator $\hat{p}=i(a^{\dagger}-a)/\sqrt{2}$, we can infer the result of applying the symmetry transformation to the bosonic field, i.e.,
$({\cal{PT}}) a ({\cal{PT}}) = -a$, $({\cal{PT}}) a^{\dagger}({\cal{PT}}) = -a^{\dagger}$ and $({\cal{PT}}) i ({\cal{PT}}) = -i$~\cite{bender2007making,BenderBook}.
We have used $[{\cal{P}},{\cal{T}}]=0$ for derivations. 
In the case of two-mode fields, these transformations of both field-mode
annihilation operators are sometimes sufficient to reveal $\cal{PT}$ symmetry 
manifesting itself as real-valued energy spectra (e.g., see Refs.~
\cite{bender2007making,BenderBook}).
Unfortunately, it is not the case of Hamiltonian~(\ref{eq:HamiltonianPT}).
One can easily check using ${\cal{P}}=\exp[i\pi(a^{\dagger}a+b^{\dagger}b)]$~\cite{bender2007making,BenderBook} that $({\cal{PT}})H_{1}({\cal{PT}}) \neq H_{1}$ because of the last two terms of $H_{1}$, which describe loss in the mode $A$ and gain in the mode $B$. The gain-loss terms are, however, crucial for the existence of EPs, because non-Hermiticity is necessary for the emergence of EPs~\cite{Mostafazadeh15,minganti2020hybridliouvillian}. Moreover, modelling losses is unavoidable as they are present in all real systems.

Therefore, to properly account for the symmetry between loss and gain we can, e.g., modify the 
space-reflection operator by multiplying it by the exchange operator $P_{\rm{S}}$ \cite{HorodeckiPRL02}, which can be interpreted as exchanging the modes
spatially (i.e., $a\leftrightarrow b$). Since there are only two modes, the exchange operator
acts in the same way as a permutation operator or a cyclic-shift operator, and thus,
we can use all these terms interchangeably here. 
A matrix representation of $P_{\rm{S}}$ is given by a perfect shuffle~\cite{Loan00}.
Hence, we use here the modified parity operator given by $\widetilde{\cal{P}}=P_{\rm{S}}{\cal{P}}$ and one can easily check that $({\cal{\widetilde{P}T}})H_{1}({\cal{\widetilde{P}T}}) = H_{1}$, so $H_{1}$ is ${\widetilde{\cal{P}}{\cal{T}}}$-symmetric. Note that $P_{\rm{S}}$ does not change any important features of the parity operator because $P_{\rm{S}}$ is also a reflection operator satisfying $P_{\rm{S}}^{2}=1$, $[P_{\rm{S}},{\cal{P}}]=0$, and $[P_{\rm{S}},{\cal{T}}]=0$. Thus, ${\widetilde{\cal{P}}{\cal{T}}}$ is also a reflection and $[\widetilde{\cal{P}},{\cal{T}}]=0$. 
Using these features of ${\widetilde{\cal{P}}{\cal{T}}}$ and applying the Cayley–Hamilton theorem, one can prove that the parameters of the characteristic equation of ${\widetilde{\cal{P}}{\cal{T}}}$-symmetric Hamiltonians are real~\cite{BenderBook}. Consequently, the eigenvalues of $H_{1}$ may only appear as complex-conjugate pairs or are real. Moreover, one can show that an eigenvalue of $H_{1}$ is real if the corresponding eigenstate of $H_{1}$ is also an eigenstate of ${\widetilde{\cal{P}}{\cal{T}}}$. If all the eigenfunctions are identical for both operators, then the entire spectrum of $H_{1}$ is real. In such a case $H_{1}$ is said to have {\emph{unbroken}} ${\widetilde{\cal{P}}{\cal{T}}}$ symmetry. Otherwise, the  ${\widetilde{\cal{P}}{\cal{T}}}$-symmetry is {\emph{broken}} and then, complex-conjugate pairs of eigenvalues appear. It is obvious that values of various system parameters determine whether the symmetry of $H_{1}$ is broken. For example, for $\kappa=0$, the entire spectrum is real because then $H_{1}$ is Hermitian. However, one can expect that large enough $\kappa$ values result in symmetry breaking. 
We can easily illustrate this feature of $H_{1}$ in the case when $\varepsilon=0$, i.e., when lasers are turned off, because the solution for such a Hamiltonian is well known~\cite{Christodoulides:2018arz}.
For $\varepsilon=0$, we can diagonalise the Hamiltonian by introducing new bosonic operators, 
which are superpositions of the old ones, and which describe eigenmodes. Eigenvalues are
then given by $n\lambda$, where $\lambda=\sqrt{g^2-\kappa^2}$ and $n$ is the difference between
the numbers of excitations in the first and second eigenmode.
\begin{figure*}[ht]
    \centering
    \includegraphics[width=1.0\linewidth]{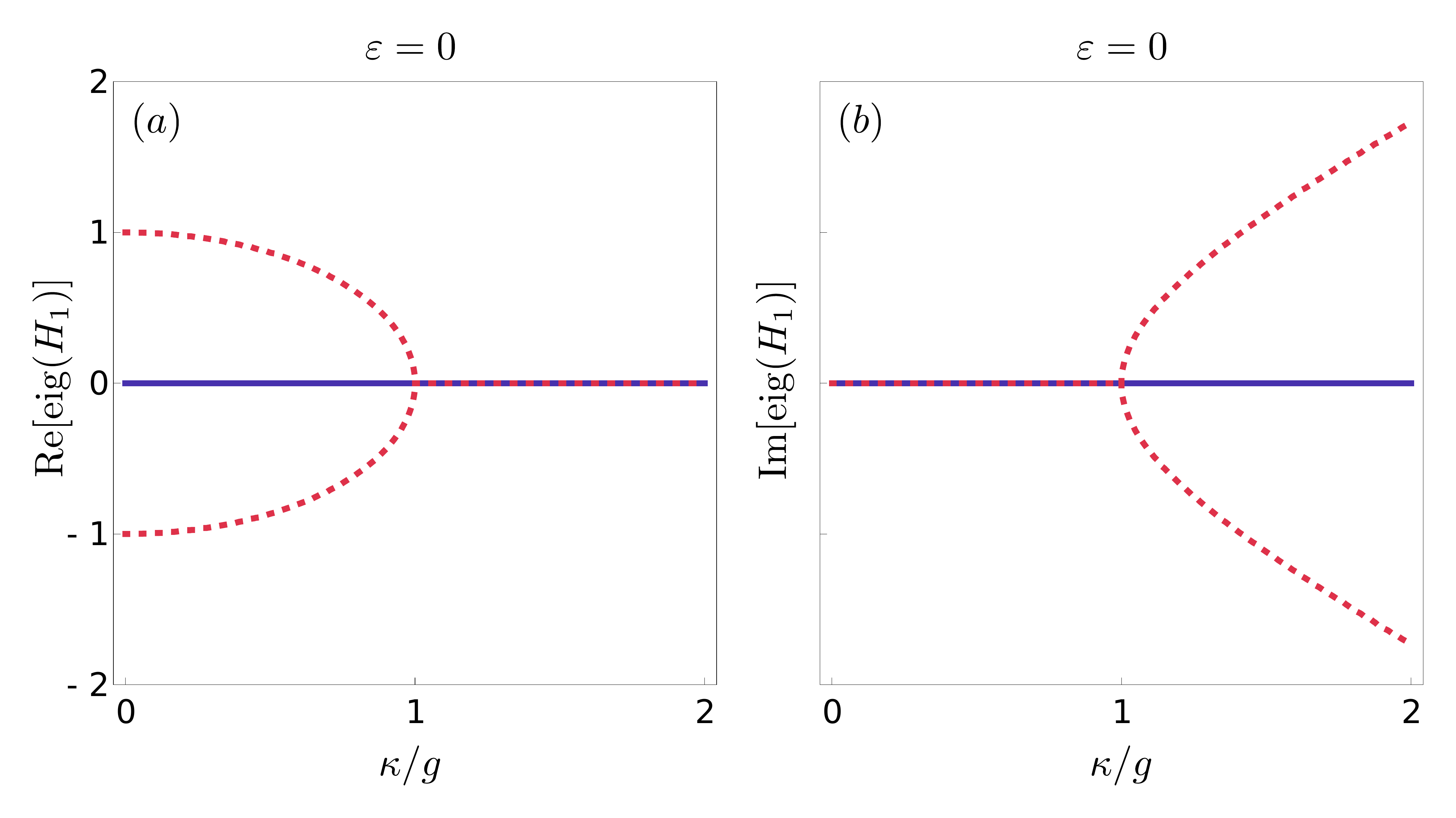}
    \caption{Spectrum of the ${\widetilde{\cal{P}}{\cal{T}}}$-symmetric Hamiltonian $H_{1}$ as a function of the gain and loss parameter $\kappa$ in units of the intercavity coupling strength $g$. Panel (a) shows the real part of the first three eigenvalues for the drive strength $\varepsilon$ = 0, whereas panel (b) shows their imaginary parts. One can observe the regions where the ${\widetilde{\mathcal{P}}{\cal{T}}}$ symmetry is unbroken, and, thus, the eigenvalues are real, and the regions where the symmetry is broken and the eigenvalues are complex-conjugate pairs (dotted curves).}
    \label{fig:spectrum_PT}
\end{figure*}
In Fig.~\ref{fig:spectrum_PT}, we plot real and imaginary parts of three eigenvalues (corresponding to $n=-1,0,1$) as a function of $\kappa$ for $g=1$.
One can see that for $\kappa<g$ all these eigenvalues are real. However, for
$\kappa>g$, one eigenvalue is real, while the two other make a complex-conjugate pair. 
One can also see that exactly at the point $\kappa=g$ a symmetry-breaking transition takes place. 
All eigenvalues have there the same real and imaginary parts --- they are identical. In such points not only the eigenvalues are the same, but also the corresponding eigenvectors are parallel, and therefore, the Hamiltonian is not diagonalisable~\cite{ozdemir2019parity,el2018non}. Such points in the parameter space are known as EPs. Since some of eigenvectors are parallel in EP, Hermitian Hamiltonians cannot exhibit any EP~\cite{Mostafazadeh15,minganti2020hybridliouvillian}. The Hamiltonian~(\ref{eq:HamiltonianPT}) is non-Hermitian, so in Fig.~\ref{fig:spectrum_PT} we can observe an EP.

Spectral singularities are of great importance because non-Hermitian systems in their vicinity can exhibit phenomena, which cannot be observed in Hermitian systems. 
It is worth mentioning that in case of two or more dimensional parameter spaces it possible to observe not single points but exceptional lines or even exceptional surfaces~\cite{ZhouOPT19,OkugawaPRB19}. Exceptional surfaces have similar properties as EPs, but are more stable and easier achievable in experiments~\cite{ZhongPRL19}.

\subsection{The effect of laser pumping on a spectral singularity}
We are going to investigate the influence of a non-zero $\varepsilon$ on the spectrum of the Hamiltonian $H_{1}$ and on its spectral singularity. For this purpose, we diagonalise $H_{1}$ by expressing it in terms of the bosonic operators $c_{\varepsilon}$, $c_{\varepsilon}^{+}$, $d_{\varepsilon}$, and $d_{\varepsilon}^{+}$ (see Methods), i.e.,
\begin{eqnarray}
H_{1}&=&\lambda\,(c_{\varepsilon}^{+} c_{\varepsilon}-d_{\varepsilon}^{+} d_{\varepsilon})
+\lambda_{0} \hat{I} \, ,
\end{eqnarray}
where $\lambda_{0}=-2 g\varepsilon^2/\lambda^2$.
Let us restrict ourselves to considering only three different eigenvalues --- one of them corresponding to the cases, in which there are equal numbers of excitations in both $c_{\varepsilon}$ and $d_{\varepsilon}$ modes:
\begin{eqnarray}
  \label{eq:E0}
E_{0}&=&-\frac{2 g\varepsilon^2}{\lambda^2} \, ,
\end{eqnarray}
and the two others corresponding to the cases, in which there is one excitation more in one mode
than in the other:
\begin{eqnarray}
  \label{eq:Epm}
E_{\pm}&=&\pm \lambda- \frac{2 g\varepsilon^2}{\lambda^2} \, .
\end{eqnarray}

\begin{figure*}[ht]
    \centering
    \includegraphics[width=1.0\linewidth]{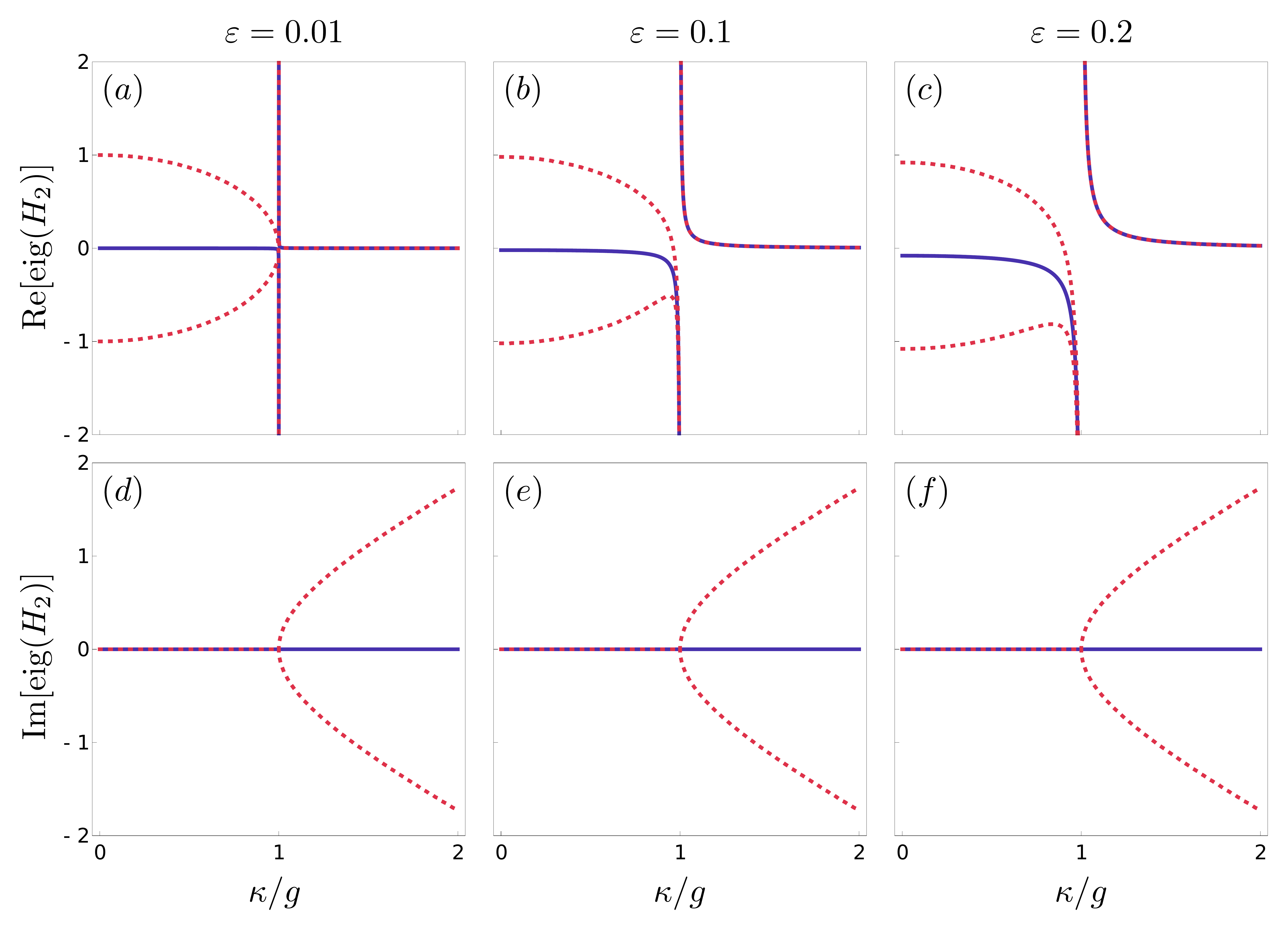}
    \caption{Spectrum of the Hamiltonian $H_{2}$ as a function of the gain and loss parameter $\kappa$ in units of the intercavity coupling strength $g$. We have plotted the real (a-c) and imaginary (d-f) parts of the eigenvalues given by Eq.~(\ref{eq:E0}) (solid curves) and by Eq.~(\ref{eq:Epm}) (dashed curves) for different values of the drive strengths: $\varepsilon$ = 0.01 (a,d), $\varepsilon$ = 0.1 (b,e) and $\varepsilon$ = 0.2 (c,f).}
    \label{fig:spectrum_RT}
\end{figure*}
As one can see from Fig.~\ref{fig:spectrum_RT}, the effect of laser pumping on the spectrum is to
shift real parts of all the eigenvalues by the same value, i.e., $\lambda_{0}$. The absolute value 
of the shift increases with the increase of $\kappa/g$ ratio. As a result we observe an unusual spectral singularity. All three eigenvalues tend to the same value in their real and imaginary parts as $\kappa\to g$, but unlike typical EP, in the case of this spectral singularity, all eigenvalues go to $-\infty$.
It is also seen from Fig.~\ref{fig:spectrum_RT} that a non-zero $\varepsilon$ does not influence
the imaginary parts of the eigenvalues.
Note that despite of the presence of this unusual spectral singularity, eigenvalues behave as we
expect from a $\cal{P}{\cal{T}}$-symmetric system. There are ranges of $\kappa$, for which eigenvalues are real, and such ranges that a complex-conjugate pair appear. Spectral singularity marks the boundary between these two ranges and in its vicinity we can observe phenomena, which cannot be observed in Hermitian systems.
For instance, one can check that we can obtain the enhancement of sensing, since 
eigenvalue splitting is given by $\Delta E=E_{+}-E_{0}\approx \sqrt{2\kappa\Delta g}$, where $\Delta g=g-\kappa$ can be interpreted as a perturbation strength.

We conclude this section by noting that the lasers pumping process does not change any typical feature of $\cal{P}{\cal{T}}$-symmetric systems except that eigenvalues tend to $\pm\infty$ while going to coalescence. To our knowledge, this kind of behaviour was not predicted yet even in other systems.

\subsection{Spectral singularities beyond the ${\widetilde{\cal{P}}\cal{T}}$ symmetry}
Remarkably, spectral singularities can be found also in non-Hermitian systems that are not $\widetilde{\cal{P}}{\cal{T}}$-symmetric~\cite{Mostafazadeh15,minganti2020hybridliouvillian}. Therefore now the question arises: can a given bosonic non-$\widetilde{\cal{P}}{\cal{T}}$-symmetric systems display any singularity and how to engineer such systems? In order to answer this question let us consider the following Hamiltonian
\begin{eqnarray}
  \label{eq:Hamiltonian_nonPT}
  H_{2}&=& g (a^{\dagger}b + b^{\dagger}a)
    +\varepsilon (a + a^{\dagger}) +\varepsilon (b + b^{\dagger})
    -i\kappa a^{\dagger}a+i\kappa b^{\dagger}b    \, .
\end{eqnarray}
It is seen that $H_{2}$ is not ${\widetilde{\cal{P}}{\cal{T}}}$-symmetric. Nevertheless, eigenvalues of $H_{2}$ are exactly the same as those of $H_{1}$ (see Methods).
Therefore it is reasonable to expect that $H_{2}$ has some other type of symmetry. In order to find this symmetry, we note that the reflection operator ${\cal{P}}=\exp[i\pi (a^{\dagger}a+b^{\dagger}b)]$ describes, in fact, the rotation of the frame by an angle $\pi$. Zhang~{\emph{et al.}} have shown that in fermionic systems the rotation-time symmetry is a generalisation of the parity-time symmetry~\cite{zhang2013non, PhysRevA.88.042108}. Therefore, we can expect that the symmetry we are looking for in our bosonic system is also the rotation-time symmetry.

Let us introduce the rotation operator given by 
\begin{equation}\label{eq:Rtheta}
{\cal{R}}=\exp[i\theta (a^{\dagger}a+b^{\dagger}b)]\, ,
\end{equation} 
where $\theta$ is an arbitrary angle and let us see whether the ${\cal{RT}}$ symmetry leads to the same conclusions as those for the ${\cal{PT}}$ symmetry. We need to apply again the permutation operator $P_{\rm{S}}$ because of the gain-loss terms. So, we define $\widetilde{\cal{R}}=P_{\rm{S}}{\cal{R}}$.

Note that ${\cal{R}}$ has different properties than ${\cal{P}}$, i.e., it is not a reflection operator, i.e., 
${\cal{R}}^{-1}\neq {\cal{R}}$, and it does not commute with the time-reflection operator ${\cal{T}}$, i.e., $[{\cal{R}},{\cal{T}}]\neq 0$. It seems to be an obstacle. Fortunately, the operator ${\widetilde{\cal{R}}{\cal{T}}}$ is a reflection operator:
\begin{eqnarray}
  \label{eq:RT2}
    ({\widetilde{\cal{R}}{\cal{T}}})^{-1} &=& {\cal{T}}^{-1} {\widetilde{\cal{R}}}^{-1}
    = {\cal{T}}(P_{\rm{S}} {\cal{R}})^{-1} = {\cal{T}} {\cal{R}}^{-1} P_{\rm{S}}
    = {\cal{T}} {\cal{R}}^{-1} {\cal{T}}^{2} P_{\rm{S}}=  {\cal{R}} P_{\rm{S}} {\cal{T}}
    =\widetilde{\cal{R}}{\cal{T}}\, .
\end{eqnarray}

Using the above property of ${\widetilde{\cal{R}}{\cal{T}}}$, we are going to show now that a non-Hermitian Hamiltonian $H$, which commutes with ${\widetilde{\cal{R}}{\cal{T}}}$, has real energy spectra, if eigenstates of ${\widetilde{\cal{R}}{\cal{T}}}$ are also the eigenstates of $H$. To this end, let us first note that, from $[H,\widetilde{\cal{R}}{\cal{T}}]=0$ and $({\widetilde{\cal{R}}{\cal{T}}})^{2} = 1$, we can infer that  the characteristic equation of $H$ is real~\cite{BenderBook}, and thus, eigenvalues of $H$ can only be real or appear as complex-conjugate pairs.

Let us demonstrate now that eigenvalues of the operator ${\widetilde{\cal{R}}{\cal{T}}}$, which we denote by $\chi$, cannot be equal to zero. In fact, they are just phase factors, i.e., the absolute value of each eigenvalue is equal to unity. The eigenvalue equation reads
\begin{equation}
  \label{eq:eig_xi}
    {\widetilde{\cal{R}}{\cal{T}}} |\Psi\rangle = \chi |\Psi\rangle \, .
\end{equation}
Multiplying Eq.~(\ref{eq:eig_xi}) on the left by ${\widetilde{\cal{R}}{\cal{T}}}$ and using the fact that $({\widetilde{\cal{R}}{\cal{T}}})^{2} = 1$ we obtain
\begin{equation}
    |\Psi\rangle = ({\widetilde{\cal{R}}{\cal{T}}}) \chi ({\widetilde{\cal{R}}{\cal{T}}})^{2} |\Psi\rangle \, .
\end{equation}
Now we can use Eqs.~(\ref{eq:eig_xi}) and (\ref{eq:RT2}), and ${\cal{T}}={\cal{T}}^{-1}$ to make further transformation:
\begin{equation}
    |\Psi\rangle = \widetilde{\cal{R}} {\cal{T}} \chi {\cal{T}} \widetilde{\cal{R}}^{-1} \chi |\Psi\rangle
\end{equation}
and, then, using ${\cal{T}} i {\cal{T}}=-i$, we finally obtain
\begin{equation}
   |\Psi\rangle = \chi^* \chi |\Psi\rangle = |\chi|^2 |\Psi\rangle \, .
\end{equation}
This concludes the proof that $\chi$ cannot be equal to zero.

Next, let us show that the eigenvalue  $E$ of a Hamiltonian $H$ is real if $|\Psi\rangle$ is an eigenstate of both operators, $H$ and ${\widetilde{\cal{R}}{\cal{T}}}$.
We begin with the time-independent Schr{\"o}dinger equation
\begin{equation}
    H |\Psi\rangle = E |\Psi\rangle \, .
\end{equation}
Multiplying it on the left by ${\widetilde{\cal{R}}{\cal{T}}}$, once more recalling $({\widetilde{\cal{R}}{\cal{T}}})^{2} = 1$ and using assumption that $|\Psi\rangle$ is also an eigenstate of ${\widetilde{\cal{R}}{\cal{T}}}$ we obtain
\begin{equation}
  \label{eq:ReE01}
   ({\widetilde{\cal{R}}{\cal{T}}}) H |\Psi\rangle = ({\widetilde{\cal{R}}{\cal{T}}}) E ({\widetilde{\cal{R}}{\cal{T}}}) \chi |\Psi\rangle \, .
\end{equation}
Since we have assumed that $H$ is ${\widetilde{\cal{R}}{\cal{T}}}$-symmetric, i.e., it commutes with ${\widetilde{\cal{R}}{\cal{T}}}$ and because ${\widetilde{\cal{R}}{\cal{T}}}=({\widetilde{\cal{R}}{\cal{T}}})^{-1}$, we can rewrite Eq.~(\ref{eq:ReE01}) in the form
\begin{equation}
   H ({\widetilde{\cal{R}}{\cal{T}}}) |\Psi\rangle = {\widetilde{\cal{R}}{\cal{T}}} E {\cal{T}} {\widetilde{\cal{R}}}^{-1} \chi |\Psi\rangle \, ,
\end{equation}
which leads us to
\begin{equation}
    E \chi |\Psi\rangle = E^* \chi |\Psi\rangle \, .
\end{equation}
Since $\chi$ is nonzero, pure phase, we deduce that $E = E^*$, which means that the eigenvalue $E$ is real.

Hence, the symmetry described by the rotation-time operator ${\widetilde{\cal{R}}{\cal{T}}}$
leads to the same conclusions as those for ${\cal{PT}}$-symmetry: (i) eigenvalues of ${\widetilde{\cal{R}}{\cal{T}}}$-symmetric Hamiltonians are real if corresponding eigenstates are the same for both ${\widetilde{\cal{R}}{\cal{T}}}$ and $H$; (ii) or eigenvalues appear as complex-conjugate pairs, when the symmetry is broken. Most importantly, we can find EPs at the boundary between the broken and unbroken ${\widetilde{\cal{R}}{\cal{T}}}$ symmetry regions in the parameter space.

We conclude this section noting that $H_{2}$ is ${\widetilde{\cal{R}}{\cal{T}}}$-symmetric
in the rotation operator~(\ref{eq:Rtheta}) with $\theta=0$.

\subsection{$\widetilde{\cal{R}}{\cal{T}}$-symmetric bosonic systems}
The ${\widetilde{\cal{R}}{\cal{T}}}$ symmetry can play an important role in searching for bosonic systems with EPs. Note that the ${\widetilde{\cal{R}}{\cal{T}}}$ symmetry includes much larger class of bosonic systems than the ${\widetilde{\cal{P}}{\cal{T}}}$ symmetry because operator ${\widetilde{\cal{P}}{\cal{T}}}$ is just one specific instance (corresponding to $\theta=\pi$) of all ${\widetilde{\cal{R}}{\cal{T}}}$ operators. To give an example of a class of ${\widetilde{\cal{R}}{\cal{T}}}$-symmetric Hamiltonians with potential EPs, let us consider a Hamiltonian, which is a generalisation of both previously analysed Hamiltonians $H_{1}$ and $H_{2}$:
\begin{eqnarray}
  \label{eq:HamiltonianRT}
  H_{3}&=&\Delta a^{\dagger}a+\Delta b^{\dagger}b 
  + g a^{\dagger}b + g^{*} b^{\dagger}a +\varepsilon (e^{i\phi} a + e^{-i\phi} a^{\dagger}) +\varepsilon (e^{i\phi} b 
    + e^{-i\phi} b^{\dagger}) \nonumber\\
    &&-i\kappa a^{\dagger}a+i\kappa b^{\dagger}b    \, .
\end{eqnarray}
The effect of an action of the ${\widetilde{\cal{R}}\cal{T}}$ operator on the Hamiltonian can be quickly inferred using the formulas:
${\widetilde{\cal{R}}{\cal{T}}} a {\widetilde{\cal{R}}{\cal{T}}} = b \exp(-i\theta)$, 
${\widetilde{\cal{R}}{\cal{T}}} b {\widetilde{\cal{R}}{\cal{T}}} = a \exp(-i\theta)$, 
${\widetilde{\cal{R}}{\cal{T}}} a^{\dagger} {\widetilde{\cal{R}}{\cal{T}}}=b^{\dagger} \exp(i\theta)$
${\widetilde{\cal{R}}{\cal{T}}} b^{\dagger} {\widetilde{\cal{R}}{\cal{T}}}=
a^{\dagger} \exp(i\theta)$ and
${\widetilde{\cal{R}}{\cal{T}}} i {\widetilde{\cal{R}}{\cal{T}}} = -i$.
Thus, using $({\widetilde{\cal{R}}{\cal{T}}})^2=1$ one can easily derive ${\widetilde{\cal{R}}{\cal{T}}} H_{3} {\widetilde{\cal{R}}{\cal{T}}}$. The result of this transformation is
\begin{eqnarray}
H_{3}^{\cal{\widetilde{R}T}} &=& \Delta b^{\dagger}b+\Delta a^{\dagger}a + g^* b^{\dagger}a 
+ g a^{\dagger}b 
+|\varepsilon| (e^{-i(\phi+\theta)} b + b^{\dagger}e^{i(\phi+\theta)}) 
+|\varepsilon| (e^{-i(\phi+\theta)} a + a^{\dagger}e^{i(\phi+\theta)})\nonumber\\
&&+i\kappa b^{\dagger}b-i\kappa a^{\dagger}a\, .
\end{eqnarray}
Hence, $H_{3}$ remains ${\widetilde{\cal{R}}{\cal{T}}}$ invariant, i.e., $H_{3}^{\cal{\widetilde{R}T}} = H_{3}$, if we choose $\theta = -2\phi$. 

It is straightforward to give expressions in their most general form, which can be a part of the Hamiltonian that governs a bosonic ${\widetilde{\cal{R}}{\cal{T}}}$-symmetric system. Each of them guarantees itself the ${\widetilde{\cal{R}}{\cal{T}}}$ invariance.
\begin{table}[ht]
\caption{Different types of ${\widetilde{\mathcal{R}}{\cal{T}}}$-invariant terms for a specific choice of ${\cal{R}}$, i.e., $\theta = -2\phi$. Parameters $n$, $m$, $j$, and $l$ are arbitrary natural numbers.}
\label{tab:RT_terms}
{\small
\begin{tabular}{lll}
\hline
 Type & \quad ${\widetilde{\cal{R}}{\cal{T}}}$ invariant terms & Examples of physical processes
 \\ \hline
 (a) & $\alpha_0 (a^{\dagger})^{n} a^{n} + H.c.$ & $n$-th order non-linearity \cite{G87,DW80,MH91} \\
 (b) & $\alpha_1 (a^{\dagger})^{m} b^{m} + H.c. $ & $m$=1 $\rightarrow$ couplers with linear exchange \cite{CB96,B93,PP00} \\
  & & $m$=2 $\rightarrow$ couplers with non-linear exchange \cite{LK11} \\
 (c) & $|\alpha_2| \big(e^{i\phi} a^{j+1} (b^{\dagger})^{j} + e^{i\phi} b^{j+1} (a^{\dagger})^{j}$ &  $j$=1 $\rightarrow$ second harmonic generation \cite{FHPW61,B08,S84,KMMSN17} \\
 &$+ H.c. \big)$ & unconventional photon blockade \cite{GeracePRA14,ZhouPRA15}\\
 &&arbitrary j $\rightarrow$ ($j$-1) harmonic generation \cite{KMMSN17}\\
 (d) & $|\alpha_3| \big(e^{i 2\phi} a^{l+2} (b^{\dagger})^{l} + e^{i 2\phi} b^{l+2} (a^{\dagger})^{l}$ & $l$=1 $\rightarrow$ third-harmonic generation \cite{B08,S84}\\
 &$+ H.c. \big)$ \\
\hline
\end{tabular}
}
\end{table}
The most relevant Hamiltonian terms having this property are presented in Tab.~\ref{tab:RT_terms}.
By utilising these ${\widetilde{\cal{R}}{\cal{T}}}$-invariant terms, one can compose a wide class of ${\widetilde{\cal{R}}{\cal{T}}}$-symmetric models, which can exhibit exceptional points.

\section{Discussion}
In summary, we have studied the ${\widetilde{\cal{R}}{\cal{T}}}$ symmetry and its applications, and conditions for its presence in a laser pumped bosonic system with losses and gain. To our knowledge, this is the first proposal of adopting this type of symmetry to quantum optics. We have shown that the ${\cal{PT}}$ symmetry is only a special case of a rotation-time symmetry. Our study of the effect of laser pumping on symmetries and symmetry phase transitions resulted in discovering a new type of spectral singularity (as shown in Fig.~\ref{fig:spectrum_RT}). This was possible by creating a new types of operators to diagonalise a Hamiltonian describing a bosonic system with a classical laser drive (see Methods). There, we have obtained real energy values without applying requirements of the $\cal{PT}$ symmetry, which turned out to be too restrictive for many types of photonic systems. We have created a versatile framework for certifying the ${\widetilde{\cal{R}}{\cal{T}}}$ symmetry based on a set of expressions that guarantee the ${\widetilde{\cal{R}}{\cal{T}}}$ invariance. We believe that this work can significantly contribute to quantum optics and we hope that can open opportunities to study new classes of systems and related physical effects.

\section{Methods}
\subsection{Diagonalisation}
Let us first show how to diagonalise the Hamiltonian $H_{2}$. For that purpose, we have to
perform the following transformations~\cite{Christodoulides:2018arz}
$[c, d]^{\rm{T}}=\boldsymbol{R}\, [a, b]^{\rm{T}}$ and 
$[c^{+}, d^{+}]^{\rm{T}}=\boldsymbol{R}\, [a^{\dagger}, b^{\dagger}]^{\rm{T}}$, where
\begin{equation}
  \label{eq:R}
\boldsymbol{R}\equiv\begin{bmatrix}
\cos\frac{\alpha}{2}&\sin\frac{\alpha}{2}\\
-\sin\frac{\alpha}{2}&\cos\frac{\alpha}{2}
\end{bmatrix}\, ,
\end{equation}
and $\alpha\in\mathbb{C}$.
For the sake of simplicity we assume here that $g$ and $\varepsilon$ are real and positive. Then it is easy to show
that $\sin{(\alpha/2)}=\sqrt{(\lambda+i\kappa)/(2\lambda)}$ and
$\cos{(\alpha/2)}=\sqrt{(\lambda-i\kappa)/(2\lambda)}$, where $\lambda=\sqrt{g^2-\kappa^2}$.
Note that $c^{+}$ ($d^{+}$) is {\emph{not}} a Hermitian conjugate of $c$ ($d$). Nevertheless,
these operators can be considered as annihilation and creation operators, because
they satisfy the canonical commutation relations $[c,c^{+}]=1$, $[d,d^{+}]=1$, $[c,d^{+}]=0$ and
$[d,c^{+}]=0$. Using these operators we can rewrite $H_{2}$ as
\begin{eqnarray}
  \label{eq:H2cd}
H_{2}&=&\lambda\,(c^{+} c-d^{+} d)+\varepsilon_{c}(c+c^{+})+\varepsilon_{d}(d+d^{+})\, ,
\end{eqnarray}
where $\varepsilon_{c}=\varepsilon \big[\cos(\alpha/2)+\sin(\alpha/2)\big]$ and 
$\varepsilon_{d}=\varepsilon \big[\cos(\alpha/2)-\sin(\alpha/2)\big]$.
One can see that we need one additional step to diagonalise the Hamiltonian.
We apply transformations given by displacement operators
$D_{c}=\exp(\varepsilon_{c}/\lambda c^{+}-\varepsilon_{c}/\lambda c)$ 
and $D_{d}=\exp(-\varepsilon_{d}/\lambda d^{+}+\varepsilon_{d}/\lambda d)$:
\begin{eqnarray}
c_{\varepsilon}&=&D^{-1}_{c} c D_{c} =c+\varepsilon_{c} \hat{I}/\lambda\, ,\nonumber\\
d_{\varepsilon}&=&D^{-1}_{d} d D_{d} =d-\varepsilon_{d} \hat{I}/\lambda\, ,\nonumber\\
c_{\varepsilon}^{+}&=&D^{-1}_{c} c^{+} D_{c} =c^{+}+\varepsilon_{c} \hat{I}/\lambda\, ,\nonumber\\
d_{\varepsilon}^{+}&=&D^{-1}_{d} d^{+} D_{d} =d^{+}-\varepsilon_{d} \hat{I}/\lambda\, ,
\end{eqnarray}
where $\hat{I}$ is the identity operator.
It is easy to check that $[c_{\varepsilon},c_{\varepsilon}^{+}]=1$, 
$[d_{\varepsilon},d_{\varepsilon}^{+}]=1$, $[c_{\varepsilon},d_{\varepsilon}^{+}]=0$ and
$[d_{\varepsilon},c_{\varepsilon}^{+}]=0$.
The Hamiltonian $H_{2}$ rewritten in terms of these operators takes the diagonal form
\begin{eqnarray}
H_{2}&=&\lambda\,(c_{\varepsilon}^{+} c_{\varepsilon}-d_{\varepsilon}^{+} d_{\varepsilon})
+\lambda_{0} \hat{I} \, ,
\end{eqnarray}
where $\lambda_{0}=-2 g\varepsilon^2/\lambda^2$.

The Hamiltonian $H_{1}$ can be diagonalised in the same way but with one additional 
step. $H_{1}$ in terms of the operators $c$, $c^{+}$, $d$, and $d^{+}$ is given by
\begin{eqnarray}
H_{1}&=&\lambda\,(c^{+} c-d^{+} d)+\varepsilon_{c}(i c-i c^{+})+\varepsilon_{d}(i d -i d^{+})\, .
\end{eqnarray}
In this additional step we introduce new operators: $i c\to c$, $-i c^{+}\to c^{+}$,
$i d\to d$, and $-i d^{+}\to d^{+}$. These new operators satisfy the same commutation relations
as original operators. In this way we transform $H_{1}$ to the form given by Eq.~(\ref{eq:H2cd}).
Then we apply transformations given by the displacement operators to obtain the same eigenvalues
as for $H_{2}$.

\section{Acknowledgements}
It is our pleasure to thank Adam Miranowicz for critical reading and helpful suggestions. This work was supported by the Polish National Science Centre (NCN) under the Maestro Grant No. DEC-2019/34/A/ST2/00081. KB also acknowledges the project No. CZ.1.05/2.1.00/19.0377 of the Ministry of Education, Youth and Sports of the Czech Republic financing the infrastructure of his workplace.


\end{document}